\documentstyle[12pt,epsfig]{article}
\begin{document}
\title{Continuous quantum measurements of a particle in a Paul trap.}

 \author{ A. Camacho  $^{\circ}$
\thanks{email: abel@abaco.uam.mx}~~and A. Camacho--Galv\'an $^{\diamond}$\thanks{email: abel@servidor.unam.mx}\\
$^{\diamond}$DEP--Facultad de Ingenier{\'\i}a,\\
 Universidad Nacional Aut\'onoma de M\'exico, \\ 
$^{\circ}$Physics Department, \\
Universidad Aut\'onoma Metropolitana-Iztapalapa. \\
P. O. Box 55-534, 09340, M\'exico, D.F., M\'exico.}
\date{}
\maketitle

\begin{abstract} 
We calculate the propagator of a particle caught in a Paul trap and subject to the continuous quantum measurement of its position. 
The probabilities of the measurement outputs, the possible trajectories of the particle, are also found. This enables us to propose 
a series of experiments that would allow to confront the predictions of one of the models that describe the interaction 
between a measured quantum system and measuring device, namely the so called  Restricted Path-Integral Formalism, 
with the experiment. 

\end{abstract}

\section{Introduction.}
The interest that the quantum theory of measurements has gendered among physicists appea\-red\- simultaneously with the inception of quantum 
theory. Between 1940 and 1980 the technology was not capable of making a repetitive measurement on a 
single quantum mechanical system. The feasible experiments in those years comprised an ensemble of single measurements 
on a large number of quantum systems, and the consequence of a measurement was the destruction of the employed quantum system. 
A deep understanding of the principles of quantum mechanics can not be achieved without a profound comprehension of quantum measurements.

Nevertheless, now we have the possibility of making repetitive measurements on a single quantum system. For instance, 
it is possible [1] to confine and observe an individual electron in a Penning trap, or we can now also have a trapped single atom and 
observe its \-in\-ter\-ac\-tion\- with a radiation field, for example by means of a laser fluorescence [2]. 
In the case of the so called Paul trap, which has led to the construction of a mass spectrometer [3], 
a ion is trapped employing a high-frequency electric quadrupole field [4]. As is known, this idea 
can also be extended to the case of neutral atoms, laser cooled and stopped atoms are confined in a magnetic quadrupole trap 
formed by two opposed, separated, coaxial current loops [5]. 

The aforementioned experiments open
the possibility of confronting the\- theo\-re\-ti\-cal\- predictions of some formalisms that pretend to describe the interaction between measuring apparatus 
and measured system against experimental results. 

One of these formalisms is the so called Restricted Path--Integral Formalism 
(RPIF) [6]. The main idea in this approach is the restriction by means of a weight functional of the integration domain of the 
path--integral that renders the \-corres\-pond\-ing\- propagator of the analyzed system, when one or more of its parameters 
are subject to a continuous measurement process. 

Let us explain this point a little bit better, and suppose that we have a particle which has a one--dimensional movement. 
The amplitude $A(q'', q')$ for this particle to move from the point $q'$ to the point $q''$ is called propagator. 
It is given by Feynman [7] 

\begin{equation}
 A(q'', q') = \int d[q]exp({i\over \hbar}S[q]),
\end{equation}
 \noindent here we must integrate over all the possible trajectories $q(t)$ and $S[q]$ is the action of the system, which is 
defined as

\begin{equation}
S[q] = \int_{t'}^{t''}dtL(q, \dot{q}).
\end{equation}

Let us now suppose that we perform a continuous measurement of the position of this particle, 
such that we obtain as result of this measurement process a certain output $a(t)$. In other words, the measurement process gives the value $a(t)$ 
for the \-coor\-di\-na\-te $q(t)$ at each time $t$, and this output has associated a certain error $\Delta a$, which is determined by the 
experimental resolution of the measuring device. The amplitude $A_{[a]}(q'', q')$ can be now thought of as a probability amplitude for the continuous measurement process to give the result $a(t)$. 
Taking the square modulus of this amplitude allows us to find the probability density for different measurement outputs.

Clearly, the integration in the Feynman path--integral should be restricted to those trajectories that match with the experimental output. 
RPIF says that this condition can be introduced by means of a weight functional   $\omega_a[q]$ [8]. 
This means that expression (1) 
becomes now under a continuous measurement process

\begin{equation}
A_a = \int d[q]\omega_a[q]exp(iS[q]).
\end{equation}

The more probable the trajectory $[q]$ is, according to the output $a$, the bigger that $\omega_a[q]$ becomes [6], 
this means that the value of $\omega_a[q]$ is approximately one for all trajectories $[q]$ that agree with the measurement 
output $a$ and it is almost 0 for those that do not match with the result of the experiment. 

Clearly, the weight functional contains 
all the information about the interaction between measuring device and measured system, and the problems that in this context appear are two:

(i) The concrete form of the weight functional $\omega_{[q]}$ depends on the measuring device [9], in other words, 
the involved experimental constructions determine these weight functionals. We must determine $\omega_{[a]}[q]$ starting from the 
knowledge that we have of the measuring device, a non trivial problem. 

(ii) If we wish an analytical expression for $A_a$, then the resulting weight functional 
must render an analytically handleable functional integral. 

The aforementioned formalism has been used to analyze several measuring processes, for instance, the measurability of an electromagnetic field 
[10], the response of a gravitational wave antenna of Weber type [6], the analysis of the measuring process of a gravitational wave in a 
laser--interferometer detector [11], and even to explain the appearence of some classical properties, like 
the concept of time, in quantum cosmology [12]. 

Even though there are already some works [13] that give the possibility of checking the predictions of RPIF against the experiment, 
it seems that we still need a better understanding of this point. 
In this context, the case of an electrically charged particle caught in a Paul trap [4] is an experiment which can be performed and in consequence 
yields a framework which may be used to check predictions of RPIF.
More precise, in this work we evaluate (using RPIF) the propagator of a harmonic oscillator with a time dependent frequency, and which is subject to the continuous measurement of its position. 
The probability of any measurement output, the possible trajectories, is also found. In other words, we apply RPIF to the case in which we measure 
continuously the position of a particle caught in a Paul trap.
\bigskip
\bigskip

\section{Propagator of a particle in a Paul trap.}
\bigskip
\bigskip

In order to obtain a confrontation with the experiment of the predictions of RPIF we consider the case of a particle caught in a Paul trap. 
It is already known that, in this case we have a harmonic oscillator having a frequency equal to $\bar{U} - \bar{V}cos(\omega t)$, 
being $\bar{U}$, $\bar{V}$, and $\omega$ constants which depend on the electric quadrupole field used to trap the 
particle. The motion equations of an electrically charged particle caught in a Paul trap are given by [4] (here we will assume that ions are injected in the $y$--direction and that there is electric field only along the $x$-- and $z$--coordinates)

\begin{equation}
\ddot{x}(t) + {e\over mr^2}[\bar{U} - \bar{V}cos(\omega t)]x(t) = 0,
\end{equation}

\begin{equation}
\ddot{z}(t) - {e\over mr^2}[\bar{U} - \bar{V}cos(\omega t)]z(t) = 0.
\end{equation}

\noindent here $e$ is the electric charge of the particle, $m$ its mass and $2r$ the distance between the electrodes that conform part of the 
experimental apparatus. The solutions to these equations are the so called Mathieu functions [14]. 

Let us now first consider the motion only along the $x$--axis. This is no restriction at all, because the complete motion can be separated in 
two independent motions. Thus, our starting point is the following Lagrangian 

\begin{equation}
L = {1\over 2}m\dot{x}^2(t) - {1\over 2}m[U - Vcos(\omega t)]x^2(t).
\end{equation}
\noindent here we have introduced the following definitions: $U = {e\over mr^2}\bar{U}$ and $V = {e\over mr^2}\bar{V}$. 
The propagator of this particle is then

\begin{equation}
U(x'', t''; x', t) = \int_{x'}^{x''}d[x(t)]exp\{{i\over \hbar}\int_{t'}^{t''}\{{1\over 2}m\dot{x}^2(t) - {1\over 2}m[U - Vcos(\omega t)]x^2(t)\}dt\}.
\end{equation}

The complete propagator will be the product of expression (7) and the\- co\-rres\-pond\-ing\- propagator along the $z$--axis, 
but for simplicity, at this point, we take only the motion along the $x$--coordinate. Later we will consider the whole motion. 
Expression (7) is valid, if and only if, no measuring process is present. Let us now suppose that we perform a continuous monitoring of the position of this particle. 
Then, according to RPIF, we must introduce a weight functional, and the difficulty here comprises the choice of this functional. 
As was mentioned above,  the concrete form of the weight functional $\omega_{[a]}[q]$ depends on the measuring device [9], thus, 
the involved experimental constructions determine these weight functionals. 
At this point we introduce an additional assumption, our choice is 

\begin{equation} 
\omega_{[a]}[x] = exp\{ -{2\over T\Delta a^2}\int_{t'}^{t''}[x(t) - a(t)]^2dt\},
\end{equation} 
\bigskip

\noindent here $T = t'' - t'$, and $\Delta a$ is the error in the position measurement, which is determined by the resolution of the measuring apparatus.

Thus, the propagator of a particle in a Paul trap and subject to the measuring of its $x$--coordinate becomes

{\setlength\arraycolsep{2pt}\begin{eqnarray}
U_{[a]}(x'', t''; x', t) = \int_{x'}^{x''}d[x]exp\{ -{2\over T\Delta a^2}\int_{t'}^{t''}[x(t) - a(t)]^2dt\}
\times\nonumber\\
exp\{{i\over \hbar}\int_{t'}^{t''}\{{1\over 2}m\dot{x}^2(t) - {1\over 2}m[U - Vcos(\omega t)]x^2(t)\}dt\}.
\end{eqnarray}} 

This propagator can be re--written as follows

{\setlength\arraycolsep{2pt}\begin{eqnarray}
U_{[a]}(x'', t''; x', t) = exp\{ -{2\over T\Delta a^2}\int_{t'}^{t''}a^2(t)dt\}\times\nonumber\\
\int_{x'}^{x''}d[x]exp\{{i\over \hbar}\int_{t'}^{t''}\{{1\over 2}m\dot{x}^2(t) - {1\over 2}m\tilde{w}^2(t)x^2(t) + F(t)x(t)\}dt\}.
\end{eqnarray}} 

Here we have that $\tilde{w}^2(t) = U - 2Vcos(\omega t) + 4{\hbar \over imT\Delta a^2}$ and $F(t) = 4{\hbar \over iT\Delta a^2}a(t)$. From now 
on we will write $\tilde{U} = U + 4{\hbar \over imT\Delta a^2}$.
In other words, the measuring process allows us to define an effective Lagrangian $L_{eff.}$ as follows

\begin{equation}
L_{eff.} = {1\over 2}m\dot{x}^2(t) - {1\over 2}\tilde{w}^2(t)x^2(t) + F(t)x(t).
\end{equation}

Let us now denote by $q(t)$ the solution to the classical equations of motion, namely, $m\ddot{q}(t) + m\tilde{w}^2(t)q(t) = F(t)$, 
and let us write 
$x(t) = q(t) + \eta(t)$, here $\eta(t') = \eta(t'') = 0$. These two last conditions have to be imposed because we have that $q(t') = x'$ and $q(t'') = x''$. 
Therefore, the propagator becomes

{\setlength\arraycolsep{2pt}\begin{eqnarray}
U_{[a]}(x'', t''; x', t) = exp\{ {i\over \hbar}S_{cl.}-{2\over T\Delta a^2}\int_{t'}^{t''}a^2(t)dt\}\times\nonumber\\
\int_{0}^{0}d[\eta(t)]exp\{{im\over 2\hbar}\int_{t'}^{t''}[\dot{\eta}^2(t) -  \tilde{w}^2(t)\eta^2(t)]dt\},
\end{eqnarray}} 

\noindent here $S_{cl.}$ denotes the corresponding classical action of the system.

But we know [15] that 

\begin{equation}
\int_{0}^{0}d[\eta(t)]exp\{{im\over 2\hbar}\int_{t'}^{t''}[\dot{\eta}^2(t) -  \tilde{w}^2(t)\eta^2(t)]dt\} = 
\sqrt{{m\over 2i\pi \hbar f(t'')f(t') \int_{t'}^{t''}f^{-2}(t)dt}},
\end{equation}

\noindent where $f(t)$ is a solution to the following differential equation 

\begin{equation}
[{d^2\over dt^2} + \tilde{w}^2(t)]\psi(t) = 0, 
\end{equation}

\noindent and satisfying also the condition $\psi(t''), \psi(t') \not =0$.

Let us now define the following dimensionless parameters $\tilde{t} = {\omega t\over 2}$, $p = {4\tilde{U}\over \omega^2}$, and 
$q= {2V\over \omega^2}$. Hence, expression (14) becomes now 
\begin{equation}
{d^2\psi(\tilde{t})\over d\tilde{t}^2} + [p  -2qcos(2\tilde{t})]\psi(\tilde{t}) = 0.
\end{equation}

In our particular case, we have that the solutions to this last equation are the so called Mathieu functions [14] (which in general are a series). 

We may choose [14] the following function as solution to (15) 

{\setlength\arraycolsep{2pt}\begin{eqnarray}
f(\tilde{t}) = cos(\tilde{t}) + {p - 1 - q\over q}cos(3\tilde{t}) + {(p - 9)(p - 1 - q) - q^2\over q^2}cos(5\tilde{t}) +\nonumber\\
{(p - 25)[(p - 9)(p - 1 - q) - q^2]\over q^3}cos(7\tilde{t}) + ...
\end{eqnarray}}

In order to calculate an approximated expression for our propagator we will truncate the Mathieu function as follows

\begin{equation}
cos(\tilde{t}) + \alpha cos(3\tilde{t}) = cos({\omega t\over 2}) + \alpha cos(3{\omega t\over 2}) , 
\end{equation}

\noindent where we have introduced the following definition $\alpha  = {p - 1 - q\over q}$. We may take without any problem more terms, 
but for the sake of brevity we consider only two.

 We may now write the propagator of a particle in a Paul trap, whose $x$--position is continuously measured
\begin{equation}
U_{[a]}(x'', t''; x', t) = exp\{ {i\over \hbar}S_{cl.}-{2\over T\Delta a^2}\int_{t'}^{t''}a^2(t)dt\}\times\nonumber\\
\end{equation}
$$
\sqrt{(3\,\alpha  -1) (3\,\alpha  ^2+2\,\alpha  -1)\frac{\omega m}{8\pi \alpha  \hbar}}
$$
$$
\times\Biggl[\left(-4\,\arctan({\frac {2\,{e^{\omega\,it''}}\alpha  +1-\alpha  }{\sqrt {3\,\alpha  ^{2
}-1+2\,\alpha  }}})+4\,\arctan({\frac {2\,{e^{\omega\,it'}}\alpha  +1-\alpha  }{\sqrt {3
\,\alpha  ^{2}-1+2\,\alpha  }}})\right)\times
$$
$$
{\frac {1}{\sqrt {3\,\alpha  ^{2}-i+2\,\alpha  }}}+{\frac {
\left (\alpha  +1\right )\left (-{e^{\omega\,it'}}+{e^{\omega\,it''}}
\right )}{\left ({e^{\omega\,it''}}+1\right )\alpha  \left ({e^{\omega\,it'}}+1\right )}}
$$
$$
+{\frac {\left ({e^{\omega\,it''+2\omega\,it'}}-2
\,{e^{2\omega\,it'}}+{e^{\omega\,it'}}-{e^{\omega\,it''}}-{e^{\omega
\,it'+2\omega\,it''}}+2\,{e^{2\omega\,it''}}\right )\alpha  ^{2}}{
\left (-\alpha  {e^{2\omega\,it''}}-{e^{\omega\,it''}}+{e^{\omega\,it''}
}\alpha  -\alpha  \right )\left (-\alpha  {e^{2\omega\,it'}}-{e^{\omega\,it'}}+{e^{\omega\,
it'}}\alpha  -\alpha  \right )}}
$$
$$
+{\frac {\left (-{e^{\omega\,it''+2\omega\,
it'}}+{e^{\omega\,it''}}+{e^{\omega\,it'+2\omega\,it''}}
-{e^{\omega\,it'}}\right )\alpha  }{\left (-\alpha  {e^{2\omega\,it''}}-{e^{\omega\,
it''}}+{e^{\omega\,it''}}\alpha  -\alpha  \right )\left (-\alpha  {e^{2\omega\,it'}}-{e^{\omega\,it'}}+{e^{\omega\,it'}}\alpha  -\alpha  \right )}}\Biggr]^{-{1\over 2}}\times
$$
$$
\times\Biggl[{\frac {\left ({e^{3/2\omega\,it'+3\omega\,it''}}+{e^{3/2\omega\,it'}
}-{e^{3/2\omega\,it''+3\omega\,it'}}-{e^{3/2\omega\,it''}}\right ){e^{
-3/2\omega\,it''-3/2\omega\,it'}}\alpha  }{2}}
\nonumber\\
$$

$$
+ {\frac {\left ({e^{3/2\omega\,it'+2\omega\,it''}}+{e^{3/2\omega\,it'+\omega\,it''}}-{e^{3/2\omega\,it''+2\omega\,it'}}-{e^{3/2\omega\,it''+\omega\,it'}}\right ){e^{-
3/2\omega\,it''-3/2\omega\,it'}}}{2}}\Biggr]^{-{1\over 2}}.
$$

\section{Probabilities and conclusions.}
\bigskip
\bigskip

The probability associated to a measurement process along the $x$--axis, which gives as measurement output for the $x(t)$--position of the 
particle the function $a(t)$ is given by the following expression

\begin{equation}
P_{[a]} = |U_{[a]}(x'', t''; x', t)|^2. 
\end{equation}

Looking at expressions (4) and (5), it is readily seen that the probability $P_{[b]}$ of obtaining a function $b(t)$ as result of the continuous 
measurement of the $z$--coordinate of our particle is given by

\begin{equation}
P_{[b]} = |U_{[b]}(z'', t''; z', t)|^2, 
\end{equation}

\noindent where the expression that corresponds to  this case has the same form as expression (18), with 
the following four modifications: (i) $\alpha$ becomes $\beta = {\hat{p} - 1 - \hat{q}\over \hat{q}}$; (ii) $\tilde U$ has to be substituted with 
$\hat{U} = -U + 4{\hbar\over imT\Delta b^2}$; (iii) $p$ is now $\hat{p} = {4\hat{U}\over \omega^2}$ and finally; (iv) $q$ becomes $\hat{q}= -{2V\over \omega^2}$.

Thus, the probability of having as a measurement output a function $a(t)$ for the $x$--coordinate and a function $b(t)$ for the $z$--coordinate 
is then given by 

\begin{equation}
P_{[a, b]} = P_{[a]}\cdot P_{[b]}. 
\end{equation}

This is the expression we were looking for, and it gives us the opportunity to confront the predictions of RPIF against the experiment.  
A concrete experimental case that yields this possibility appears in connection with the coordinate monitoring of an individual barium ion [2]. 
Here, one of the possible choices, for the set of involved experimental parameters, is the following one [4, 17]: $m = 2.28\times 10^{-25}$ kg, $\Delta a = 2 \mu$m, $r = 8$ mm, $\bar U = 10$ volts, 
$\bar V = 10^2$ volts, $\omega = 2$ MHz and finally, the measuring process lasts $T = 30$ seconds. These parameters mean that in expression (18) we must introduce the identification $\alpha = -(2.62 + i2.81\times 10^{-11})$. 
The corresponding definition in the case of movement along the $z$--axis is $\beta = -(1.02 +i2.81\times 10^{-11})$.

To conclude, we must mention that we have used expression (8) as the description for the interaction between measuring 
device and the particle's coordinate. Of course, that we do not expect to have a so simple expression for this interaction. 
Therefore, we must consider (8) as an approximation of the corresponding interaction, and thus we must take (18) also as 
approximated result of the correct one. 

Nevertheless, there are three reasons that justify the introduction of expression (8):

(i) In the first case when RPIF was applied, Heaveside functions were used to form the weight functional [8]. 
Afterwards, the same system, but with the introduction of a gaussian weight functional, was analyzed [16]. Both calculations gave the same results up to the order of magnitude.

(ii) We would like to obtain an analytical expression for the probability (in order to have, at least as a first approximation, theoretical predictions 
that could be confronted with some proposed experiments), and our weight functional renders a gaussian 
path--integral, which can be easily calculated.

(iii) It has already been pointed out that the concrete form of the weight factor depends on the chosen experimental realization for the measurement process. 
Nevertheless, if the measurement process is realized as a series of repeated short interactions of the measuring device with a subsidiary system 
(meter), the gaussian form of the weight functional will arise automatically [18]. Thus, it is preferable, not only \-ma\-the\-ma\-ti\-cal\-ly\- but also physically, to choose this gaussian as weight factor.

The first argument not only justifies our choice, expression (8), 
but at the same time tells us that our result could be correct, in some experimental situations, only up to the order of magnitude of the considered effects. The third remark means 
that our choice (a gaussian weight functional) renders, under certain experimental conditions, the exact description of the interaction between measuring device and measured system. 
To resume, our choice is in some particular experimental constructions the exact weight functional and in the remaining cases it could 
give us, up to the order of magnitude of the considered effects, the correct results. 

Of course 
that a more accurate description of this interaction, measuring device--measured system, is needed, but we should then take expression (18) 
only as a first approximation of our problem, which could allow us to confront the order of magnitude of the predictions of RPIF 
against the experiment. In order to obtain a more realistic picture we must analyze how the weight functional is defined by the 
measuring apparatus.
\bigskip
\bigskip
\bigskip

\Large{\bf Acknowledgments.}\normalsize
\bigskip

This work was partially supported by CONACYT Grant No. 3544--E9311. A. C. would also like to thank A. Mac{\'\i}as for the fruitful discussions on the subject 
and M. Mensky,  not only for mentioning his results before their publication but also for contributing with many suggestions during the preparation of this paper.

\end{document}